\documentclass[conference]{IEEEtran}
\IEEEoverridecommandlockouts
\usepackage{cite}
\usepackage{amsmath,amssymb,amsfonts}
\usepackage{graphicx}
\usepackage{textcomp}
\usepackage{xcolor}
\def\BibTeX{{\rm B\kern-.05em{\sc i\kern-.025em b}\kern-.08em
    T\kern-.1667em\lower.7ex\hbox{E}\kern-.125emX}}

\usepackage[inline]{enumitem}

\usepackage{pdfpages}
\usepackage{float}

\usepackage{filecontents}

\usepackage{pgf,tikz}
\usepackage{pgfplots}
    \pgfplotsset{compat=1.5.1}
    \usepgfplotslibrary{
        groupplots,
    }
    \pgfplotsset{
        compat=1.5.1,
        my style/.style={
            width=8.25cm,height=6.6cm,
            xlabel=number of cores,
            ylabel=elapsed[s],
        },
        my legend style/.style={
            legend entries={
                GPI2,
                intel-14.0.4.211-I\_MPI\_ADJUST\_ALLGATHERV\_1,
                intel-14.0.4.211-I\_MPI\_ADJUST\_ALLGATHERV\_2,
                intel-14.0.4.211-I\_MPI\_ADJUST\_ALLGATHERV\_3,
                intel-14.0.4.211-I\_MPI\_ADJUST\_ALLGATHERV\_4
            },
            legend style={
            	font=\fontsize{7.5}{5}\selectfont,
                at={([yshift=2pt]0,1)},
                anchor=south west,
            },
            legend columns=1,
        },
        cycle multiindex* list={
            blue!75!black,
            red!75!black,
            green!75!black,
            violet!75!black,
            cyan!75!black
                \nextlist
            mark=*,
            mark=square,
            mark=triangle,
            mark=triangle*,
            mark=triangle*
                \nextlist
        },
    }
\usepackage{mathrsfs}
\usetikzlibrary{arrows}
\usepackage{graphicx}
\usepackage{adjustbox}
\usepackage[small,bf]{caption}
\usepackage{subcaption}
\usepackage{color}  
\usepackage{xcolor}
    \definecolor{ListingsKeywordColor}{rgb}{0,0,0.4}
    \definecolor{ListingsIdentifierColor}{rgb}{0,0.5,0}
    \definecolor{ListingsCommentColor}{rgb}{0.4,0.4,0.4}
    \definecolor{ListingsStringColor}{rgb}{0.6000,0.3333,0.7333}
    \definecolor{ListingsRuleSepColor}{rgb}{0,0,0}
    \definecolor{ListingsEmphColor}{rgb}{0,0.6667,0.6667}
    \definecolor{ListingsBreakSymbolColor}{rgb}{0.780,0.082,0.522}
    \definecolor{LinkColor}{rgb}{0,0,0.5}
    \definecolor{UnitColor}{rgb}{0,0,0}
    \definecolor{MathsVectorColor}{rgb}{0,0,0}
    \definecolor{MathsMatrixColor}{rgb}{0,0,0}
    \definecolor{MyGreen}{HTML}{228B22}
    \definecolor{MyBlue}{HTML}{0000FF}
    \colorlet{MatrixElementsLight}{gray!40!white}
    \colorlet{MatrixElementsDark}{gray!80}
    \colorlet{MyGreenLight}{MyGreen!40!white}
    \colorlet{MyGreenDark}{MyGreen!80}    
    \colorlet{MyBlueLight}{MyBlue!40!white}
    \colorlet{MyBlueDark}{MyBlue!80}    
    \colorlet{MyRedLight}{red!20!white}
    \colorlet{MyRedDark}{red!60}    
    
\usetikzlibrary{
  external,
  arrows,
  positioning,
  decorations.pathmorphing,
  3d
}
\tikzsetexternalprefix{imgs/tikz/}
\tikzset{
    external/export=false,
    >=stealth',
    punkt/.style={
           rectangle,
           rounded corners,
           draw=black, very thick,
           text width=6.5em,
           minimum height=2em,
           text centered},
    pil/.style={
           ->,
           semithick,
           shorten <= 0pt,
           shorten >= 0pt,},
    pild/.style={
           ->,
           thick,
           shorten <= 0pt,
           shorten >= 0pt,}
}

\pgfkeys{
    /tikz/external/mode=list and make
}
\pgfplotsset{
  xtick scale label code/.code={$\times 10^{#1}$}
}
\pgfplotsset{
ytick scale label code/.code={$\times 10^{#1}$}
}
\pgfplotsset{
  invoke before crossref tikzpicture={\tikzexternaldisable},
  invoke after crossref tikzpicture={\tikzexternalenable},
}
\usepackage{multirow}
\usepackage{multicol}

\usepackage{listings}
\usepackage{url}
\usepackage[final,bookmarks=false]{hyperref}
\hypersetup{
   colorlinks=false,         
   urlcolor=LinkColor,    
   filecolor=LinkColor,  
   linkcolor=LinkColor,  
   menucolor=LinkColor,
   citecolor=LinkColor,
   raiselinks=true,       
   breaklinks,              
   verbose,
   hyperindex=true,         
   linktocpage=true,        
   hyperfootnotes=false,     
   bookmarks=true,          
   bookmarksopenlevel=1,    
   bookmarksopen=false,      
   bookmarksnumbered=true,  
   bookmarkstype=toc,       
   plainpages=false,        
   pageanchor=true,         
   pdftitle={Hierarchical Approach for Deriving a Reproducible LU factorization on GPUs}, 
   pdfauthor={Roman Iakymchuk},            
   pdfcreator={LaTeX, hyperref, KOMA-Script},
   pdfstartview=Fit,       
   pdffitwindow=true,
   pdfpagemode=UseOutlines, 
   pdfpagelabels=true,      
}

\usepackage{float}
\usepackage[ruled,norelsize]{algorithm2e}
\makeatletter
\newcommand{\removelatexerror}{\let\@latex@error\@gobble}
\makeatother

\usepackage{cleveref}
\Crefname{algorithm}{Algorithm}{Algorithms}
\usepackage{rotating}

\usepackage[group-separator={,}]{siunitx}
\usepackage{comment}

\newlength{\figwidth}
\newlength{\subfigwidth}
\usepackage{filecontents} 

\begin{document}
\bstctlcite{IEEEexample:BSTcontrol}

\title{Efficient and Eventually Consistent Collective Operations}

\author{\IEEEauthorblockN{Roman Iakymchuk\IEEEauthorrefmark{1}\IEEEauthorrefmark{2}, Am\^andio Faustino\IEEEauthorrefmark{3}, Andrew Emerson\IEEEauthorrefmark{4}, Jo\~ao Barreto\IEEEauthorrefmark{3}, Valeria Bartsch\IEEEauthorrefmark{1},\\
Rodrigo  Rodrigues\IEEEauthorrefmark{3}, 
Jos\'e C. Monteiro\IEEEauthorrefmark{3}}
\IEEEauthorblockA{\IEEEauthorrefmark{1} Fraunhofer ITWM, 67663 Kaiserslautern, Germany\\
Emails: \{roman.iakymchuk,valeria.bartsch\}@itwm.fraunhofer.de}
\IEEEauthorblockA{\IEEEauthorrefmark{2}Sorbonne Universit\'e, 75252 Paris, France}
\IEEEauthorblockA{\IEEEauthorrefmark{3}INESC-ID \& IST (ULisboa), 1000-029 Lisboa, Portugal\\ 
Email: \{amandio.faustino,joao.barreto,rodrigo.miragaia.rodrigues\}@tecnico.ulisboa.pt, jcm@inesc-id.pt}
\IEEEauthorblockA{\IEEEauthorrefmark{4}CINECA, 40033 Casalecchio di Reno, Italy\\ 
Email: a.emerson@cineca.it}
}

\maketitle
 
\begin{abstract} 

Collective operations are common features of parallel programming models that are frequently used in High-Performance (HPC) and machine/ deep learning (ML/ DL) applications. In strong scaling scenarios, collective operations can negatively impact the overall application performance: with the increase in core count, the load per rank decreases, while the time spent in collective operations increases logarithmically. 

In this article, we propose a design for eventually consistent collectives suitable for ML/ DL computations by reducing communication in Broadcast and Reduce, as well as by exploring the Stale Synchronous Parallel (SSP) synchronization model for the Allreduce collective. Moreover, we also enrich the GASPI ecosystem with frequently used classic/ consistent collective operations -- such as Allreduce for large messages and AlltoAll used in an HPC code. Our implementations show promising preliminary results with significant improvements, especially for Allreduce and AlltoAll, compared to the vendor-provided MPI alternatives.
\end{abstract}

\begin{IEEEkeywords}
 Collectives,
 Allreduce,
 AlltoAll,
 Stale Synchronous Parallel,
 GASPI.
\end{IEEEkeywords}

\section{Introduction}
The {\em Global Address Space Programming Interface (GASPI)}~\cite{gaspi15} programming model has proven to be an interesting alternative to the Message Passing Interface (MPI) model, in large part due to its open source implementation, {\em GPI-2}\footnote{GPI-2's repository: \url{https://github.com/cc-hpc-itwm/GPI-2}} and support for modern hardware architectures.

In this paper, our focus is on collective operations involving all the processes, which are frequently used in distributed computations from HPC to distributed machine and deep learning (ML/ DL) computations. In particular, we make a preliminary exploration of the idea that collective operations can be designed and implemented with GASPI such that a globally consistent view is dropped. The concept comes from the distributed computing community, where early work on mobile computing proposed the notion of eventual consistency~\cite{bayou}, and has been adopted by the GRID computing community where globally distributed databases allow for reads to return potentially stale data. As the concept of {\em eventually consistent data} is ported to HPC collectives, it brings the potential to be used in many application domains, such as machine learning (ML).

In this particular domain, most ML algorithms can be classified as iterative and convergent. These algorithms start with an initial guess of a model and then improve it across several iterations until converging to a solution. 
In a typical distributed implementation of ML algorithms, several workers compute adjustments in parallel to the same model. Hence, workers are usually required to work on the same iteration, which leads to several synchronization points. However, due to the convergent nature of these algorithms, consistency can be dropped to a certain extent without jeopardizing the result. This can be achieved by allowing workers to compute less accurate adjustments using stale data from different previous iterations \cite{ssp14, SSP_W_PS}. This reduces the required synchronization, thus allowing workers to go through iterations faster. 

In this paper, we make initial contributions in exploring the space of {\em eventually consistent collectives} by investigating a Stale Synchronous Parallel (SSP) synchronization model~\cite{ssp14}, which allows the workers to compute iterations using bounded stale data. In the SSP model, the worker can receive updates while performing computation on stale data and, thus, seamlessly overlap communication and computation. We verify this implementation on an example of Matrix Factorization trained with Stochastic Gradient Descent. Additionally, we consider a possibility to drop a certain part of the data that is below a user-defined threshold in a collective. This allows the application to proceed with computations upon arrival of a part of data instead of the full amount. 

As a final contribution, we also provide {\em classic/ consistent asynchronous} variants of {\em Allreduce} suitable for large messages, such as those exchanged in ML/ DL computations, as well as in {\em AlltoAll}, which is a time-consuming operation within the Quantum Espresso~\cite{QE-2009} application. With this effort, we aim to extend and enhance the current (limited) set of collectives in GPI-2 to provide developers/ users with a {\em library of collectives} in order to facilitate their work.  

%

This article is structured as follows: \Cref{sec:gaspi} introduces the asynchronous one-sided communication in GASPI. \Cref{sec:eventcol} reports on our initial work on eventually consistent collectives, while \Cref{sec:classic} presents customized consistent collectives. \Cref{sec:results} reports  performance results. Finally, \Cref{sec:relatedworks} reviews related work, and \Cref{sec:conclusions} draws conclusions. 

\section{GASPI's asynchronous communication model}
\label{sec:gaspi}
The GASPI standard promotes the use of {\em one-sided communication}, where one side, the initiator, has all the relevant information for performing the data movement. The benefit of this is decoupling the data movement from the synchronization between processes. It enables the processes to put or get data from remote memory, without engaging the corresponding remote process, or having a synchronization point for every communication request. However, some form of synchronization is still needed in order to allow the remote process to be notified upon the completion of an operation.
In addition, GASPI provides what are known as {\em weak synchronization} primitives, which update a {\em notification} on the remote side. The notification semantics is complemented with routines that wait for the update of a single or a set of notifications. Note that similar weak synchronization primitives might also appear in the upcoming MPI-4 standard~\cite{belli2015notified}.
GASPI allows for a thread-safe handling of notifications, providing an atomic function for resetting a local notification. The notification procedures are one-sided and only involve the local process.

\begin{table*}[!t]
\caption{Communication scheme in GASPI on example of {\tt gaspi\_write\_notify} with the producer-consumer roles.}
\label{tbl:gaspi:write}
\centering
\begin{tabular}{|l|l|l|l|}
\hline
description/ usage & producer & consumer & description/ usage\\
\hline\hline
setup phase & allocate resource & allocate resources & setup phase\\
& exchange meta info & exchange meta info & \\\hline
check before communication &  & {\tt gaspi\_notify} & start as soon as the \\
 & {\tt gaspi\_wait\_some} & & receive buffer can be overwritten\\\hline
start as soon as the  & {\tt gaspi\_write\_notify} &  & check before we want to work on  \\
source buffer is filled & & {\tt gaspi\_wait\_some} & the receive buffer or wait until its filled\\\hline
check whether or wait until the &  & {\tt gaspi\_notify} & send as soon as the data arrived \\
source buffer can be overwritten & {\tt gaspi\_wait\_some} & & (acknowledgment) \\\hline
shut down phase & release resources & release resources & shut down phase\\
\hline
\end{tabular}
\end{table*}
A communication strategy for parallel applications can be as important as a numerical scheme. Usually, one tries to fit a communication strategy into a given numerical scheme by taking into account the algorithmic structure and, possibly, a spectrum of target platforms. Ultimately, one wants to hide communication (performed in the background) by overlapping computation and communication. This is the underlying principle behind GASPI -- write as early as possible and check for the arrival of the data, i.e., the notification, as late as possible (right before the data is to be used).

\begin{figure}[!ht]
  \centering
  \includegraphics[width=0.9\columnwidth]{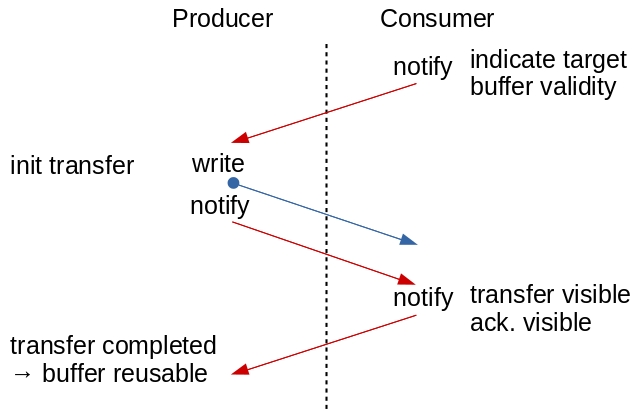}
  \caption{Generic communication scheme in GASPI on the example of {\tt gaspi\_write}.}
  \label{fig:gaspi:write}
\end{figure}
\Cref{fig:gaspi:write} outlines the communication pattern in GASPI with the help of an example of {\tt gaspi\_write}, while \Cref{tbl:gaspi:write} provides a more detailed description of the communication between producer and consumer (receiver). 
In this scenario, the consumer indicates the target buffer validity before the write begins. Then, the producer
commences with the write of the data and issues a notification; these two can be combined into one by {\tt gaspi\_write\_notify}. On the receiver side, GASPI guarantees that
data is locally available whenever the corresponding notification becomes locally visible (`transfer
visible'). This mechanism enables fine-grained (request based) asynchronous dataflow implementations
as depicted in~\Cref{fig:gaspi:write}. Finally, once the data arrives, the consumer acknowledges completion of the data
transfer; hence, the producer can reuse the buffer.

On the producer side, the {\tt gaspi\_write\_notify} --  which performs the actual communication -- is only allowed to be
invoked if the first {\tt gaspi\_wait\_some} -- which waits for the initial notification from the consumer that indicates the target buffer validity -- returns successfully.
Therefore, the two calls can be combined together into a single operation, which would be invoked as
soon as the source buffer is filled. Similarly, the {\tt gaspi\_wait\_some} and the acknowledgment of completion on the
consumer side can be combined into one operation which would be invoked before we want to work on
the receive buffer. This reduces communication to two operations on both producer and consumer sides.

\section{Eventually consistent collectives}
\label{sec:eventcol}

We envision to design eventually consistent collectives by allowing the collective (the fastest processes) to not wait for the exchange of the most recent updates. Our motivation comes from the nature of many ML/ DL algorithms that can be classified as iterative and convergent. These algorithms start with an initial guess of a model and then improve it across several iterations until converging to a solution. Their convergent nature allows these algorithms to work with data that can be a bounded number of iterations out of date (referred to as the allowed slack). This tolerance to staleness is explored in the Stale Synchronous Parallel (SSP)~\cite{ssp14} synchronization model. By allowing processes to compute iterations using bounded stale data, they can be receiving updates while performing useful computation on stale data, and thus, seamlessly overlap communication with computation. We incorporate this concept into a variant of the Allreduce collective, see~\Cref{sec:allreduce_ssp}.

Another possibility is to mimic eventually consistent collectives as follows. We can specify a termination criterion before the execution of collective, and then, in the call to collective, we can specify the predefined threshold as an input parameter. Hence, based on this threshold, every process or its parent can locally decide whether to stay silent or to engage, as well as how much to contribute. 
Consequently, the result (with respect to the threshold) can be obtained through the output status variable or at the receiver buffer after execution. This requires adding one or two additional parameters to collectives, namely threshold and status. 
 We demonstrate this idea through examples of the Broadcast and Reduce collectives, leveraging on GASPI's API (see~\Cref{sec:broadred}). 

\subsection{Allreduce with SSP}
\label{sec:allreduce_ssp}

In this section we propose an Allreduce collective following the SSP model~\cite{ssp14}. We refer to this new collective as {\tt allreduce\_SSP}. 
It is presented in 
 \Cref{hypercube-ssp-algorithm}. 
 
We used a standard hypercube allreduce as a starting point and adapted it to the SSP model. Like most allreduce algorithms, the hypercube allreduce executes in several steps (for loop in~\Cref{hypercube-ssp-algorithm}), where, in each step, communicating processes send and then wait to receive a fresh contribution, (so that these two contributions can be reduced), before continuing to the next step.
The distinguishing characteristic of the SSP model is that, instead of  waiting for fresh contributions from all processes, each process only waits until local data contains contributions from all processes made at most {\tt slack} iterations ago (lines 7-11 in~\Cref{hypercube-ssp-algorithm}). For example, if the process is in iteration 5 and allows {\tt slack} to be 1, the collective can return after using contributions from other processes that were computed in the current iteration, 5, but also from the previous iteration, 4. 
Note that, to be able to determine the age (or iteration number) of the result of reducing contributions from different iterations, the algorithm relies on a logical clock. This clock value initially corresponds to the iteration in which the contribution was computed. Then, when two contributions are reduced together, the result of that reduction is associated with the minimum clock of both contributions. For instance, if a contribution with clock 2 is reduced with another from clock 3, the reduction result would be associated with clock 2.


\begin{figure*}[htbp]
	\centering
	\includegraphics[width=\textwidth]{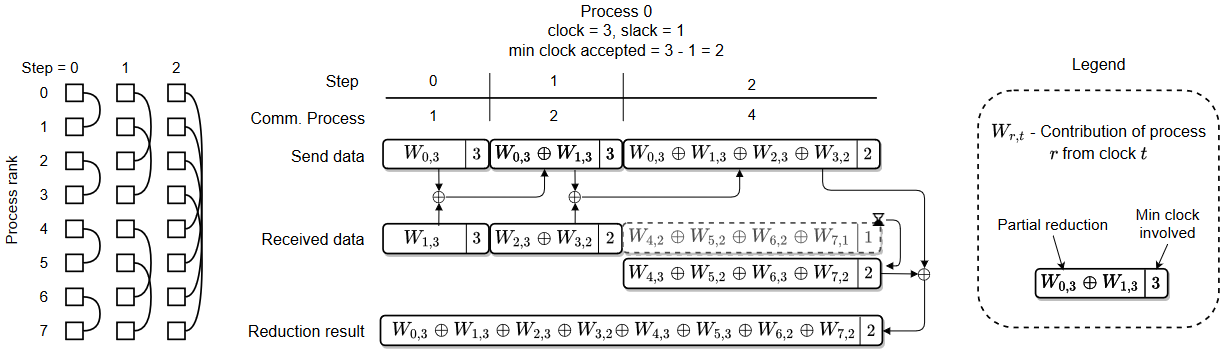}
	\caption{On the left hand side of the figure we see the communication pattern on a hypercube using 8 processes, and in the middle part, we have an example of the adaptation of the hypercube to use SSP.}
	\label{fig:hypercube-ssp-example}
\end{figure*}

Intuitively, to adapt a hypercube allreduce algorithm to support SSP, in {\tt allreduce\_SSP} processes remember the last contributions received at each step, and, provided that these contributions are not too stale, use them instead of waiting for fresh contributions.
To accomplish that, we begin by identifying the contributions received at each step by a given process and reserve dedicated memory to receive the contributions from each of the steps; then, given we are using one-sided communication, processes send updated contributions by writing them in the dedicated memory of the process requiring those contributions, thus overwriting their previously sent contributions; finally, whenever a process wants to reuse the last contributions received for a given step, it will simply read from the dedicated local memory for that step. In ~\Cref{hypercube-ssp-algorithm} we refer to this memory as \texttt{rcv\_data\_vec}, a vector with memory dedicated for each step.



\autoref{fig:hypercube-ssp-example} shows the communication pattern for the hypercube algorithm using 8 processes. The grid on the left of the figure is composed of several squares, each representing a process. The y-axis of the grid corresponds to the rank of the process, and the x-axis to the step of the algorithm. At each step, processes connected by an edge in the figure exchange contributions and reduce the received contribution with the contribution they have sent, resulting in a partial reduction, which is to be communicated in the next step. We refer to this reduction as partial as it does not contain all contributions. After following this procedure for enough steps
($\lceil\log(P)\rceil$, where $P$ is the number of processors),
we obtain the final reduction result. 

From this communication pattern, we see that process 0 receives the contribution from process 1 at step 0, a partial reduction from process 2 at step 1, and finally a partial reduction from process 4 at step 2. To implement this, we leverage the predictability of data transfer at each step to create a dedicated memory to receive the data that each process expects to receive, read from this memory, and repeat this process for the several steps in sequence. Before reading the data, we only need to wait for fresh updates if the latest contribution received is too stale. Otherwise, we use the latest contribution received. 
\begin{algorithm}[htbp]
\LinesNumbered
\SetKwInOut{Input}{Input}
\SetKwInOut{Output}{Output}
\DontPrintSemicolon
\caption{allreduce\_ssp}
\label{hypercube-ssp-algorithm}
\tcp{Hypercube with $d$ dimensions}
\Input{new\_contribution, slack}
\Output{reduction\_result}
\BlankLine
    $clock \longleftarrow clock + 1$\;
    $min\_clock\_accepted = clock - slack$\;
    \BlankLine
    $part\_red \longleftarrow new\_contribution$\;
    \BlankLine
    \For{$0 \leq k < d$} {
        $comm\_proc \longleftarrow get\_comm\_process(k)$\; 
        \BlankLine
        \tcp{Send partial reduction}
        $send(part\_red, clock, comm\_proc)$\;
        $rcv\_data \longleftarrow rcv\_data\_vec[k]$\;
        \BlankLine
        \tcp{Wait if rcv data is too stale}
        \If{$rcv\_data.clock < min\_clock\_accepted$} {
            $wait\_for\_update(k)$\;
            $rcv\_data \longleftarrow rcv\_data\_vec[k]$\;
        }
        \BlankLine
        \tcp{Reduce sent with received data}
        $part\_red \longleftarrow reduce(part\_red, rcv\_data)$\;
    }
    \BlankLine
    $reduction\_result \longleftarrow part\_red$
\end{algorithm}
 
To help in the explanation of the {\tt allreduce\_ssp} algorithm, we walk through an example of its execution, which corresponds to the diagram in the middle of~\Cref{fig:hypercube-ssp-example}.

In this example, we are zooming at process 0, currently with clock 3 in a scenario where slack is set to 1. The fact that slack is equal to 1 means that the process can use data from the current clock, in this case, clock 3, but it can also use data from the previous clock, clock 2. In the first step, the process sees that it already received data from the producer and uses it to move on to the second step. At the second step, the process now finds received data that is stale at clock 2, but still fresh enough to be used in order to move on to the next step. Once it reaches the last step, it finds that the current received data is too stale too be used. In this case, and only in this case, the process waits until receiving a new update for the current step.

\subsection{Broadcast and Reduce}
\label{sec:broadred}
\begin{figure}[!ht]
  \centering
  \includegraphics[width=0.3\textwidth]{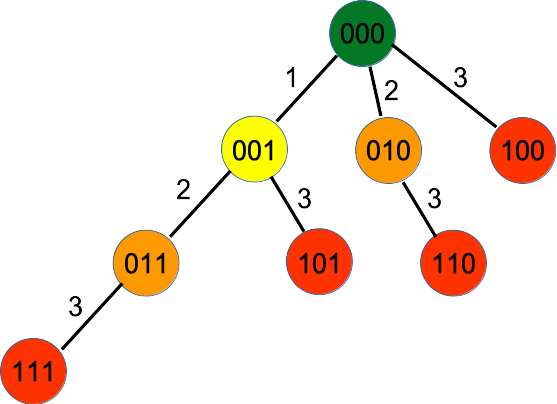}
  \caption{Broadcast using the binomial spanning tree with eight nodes; communication stages are marked on edges with colored nodes.}
  \label{fig:bst}
\end{figure}

To develop an eventually consistent Broadcast, we can rely on P-1 {\tt gaspi\_write\_notify} calls, where P is the number of processes, from the root or implement a classic binomial spanning tree (BST, see~\Cref{fig:bst})~\cite{tzeng1998fast}, specifying a certain pre-defined percentage of data to be communicated as the threshold parameter in {\tt gaspi\_bcast}. 
The BST is a common algorithm, which uses a binomial tree structure in order to reduce the network contention. 
The BST implementation of {\tt gaspi\_bcast} is rather straightforward since the data is written from root/ parent to its children. 
The algorithm begins with determining a parent and children for each process: 
rank $0$ is the root of the tree; the children of the process with rank $p_0$ are those with rank $p_0 + 2^i$, where $log(p_0) \leq i \leq\lceil log(P)\rceil$. 
Then, we apply the scheme in~\Cref{fig:gaspi:write} (a fraction of data is communicated), but only acknowledge the data transfer from the outer nodes to  their parents. 
The collective is considered complete when the outer nodes receive data.

In case of Reduce, which is the inverse of Broadcast, each child process waits for a notification from its parent indicating that the data can be sent (see~\Cref{fig:gaspi:write}). This is crucial to avoid barriers and to relax the synchronization in case of multiple children writing to the parent’s receive buffer simultaneously. The data is written to the segment on the parent’s side and reduced, and the child is also acknowledged on the completed data write. The collective continues until the root is reached and its children have contributed with their parts. 
 
With Broadcast, we can mostly rely on sending less data due to the fact that all processes, but one, should receive the full amount or a fraction of data. With Reduce, we can have two strategies: one is the same as for Broadcast; the other is to still send the full amount of data but depending on the pre-defined threshold eliminate some processes. When we look on the binomial tree, \Cref{fig:bst}, which is not balanced, we can see that it works in stages doubling the number of involved processes on each stage. Thus, we exclude some processes depending on their id and/ or the stage id, ensuring involvement of at least the threshold amount of processes. Alternatively, we can follow the deepest path on the left hand side of the tree. A disadvantage of this approach can be the existence of a varying significance of the data, which in some scenarios can be on the eliminated nodes. This can be enhanced by adding weights to the data, but we omit this in our initial study.

\section{Consistent collectives}
\label{sec:classic}
We also propose to construct consistent collectives -- such as Allreduce and AlltoAll -- by relying on GASPI's API, instead of low level APIs like ibverbs. 
 Furthermore, we outline a strategy to make Allreduce eventually consistent as well. 

\subsection{Allreduce}
Since Allreduce is primarily used for large messages (from several kilobytes to hundreds of megabytes) in ML/ DL applications, our algorithm of choice is the segment pipelined ring algorithm, as it is suitable for large message sizes and promotes communication with only two neighbors. This algorithm aims to saturate bandwidth and, hence, reach high performance. 

\begin{figure}[!ht]
\centering
\includegraphics[width=.9\columnwidth]{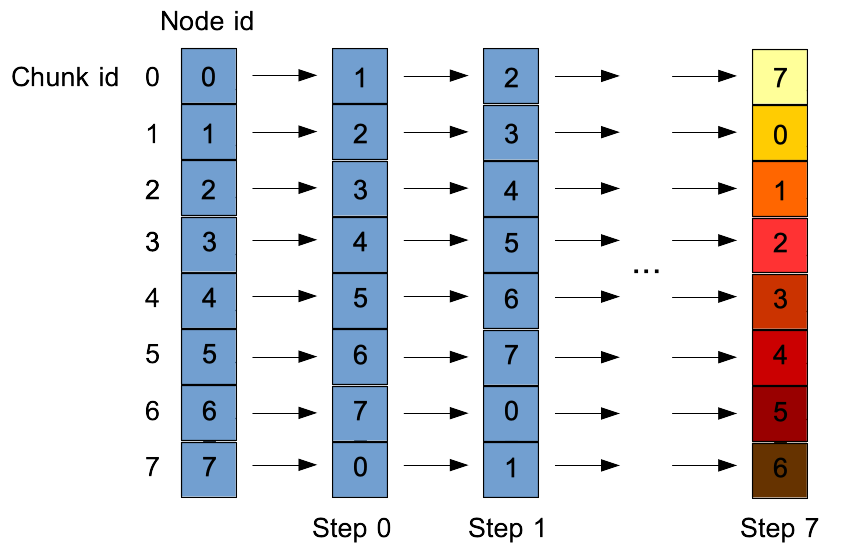}
\caption{Segmented pipelined ring Allreduce: Scatter-Reduce stage where each node has a complete partial result.}
\label{fig:allreduce_reduce}
\end{figure}
\begin{figure}[!ht]
\centering
\includegraphics[width=0.73\columnwidth]{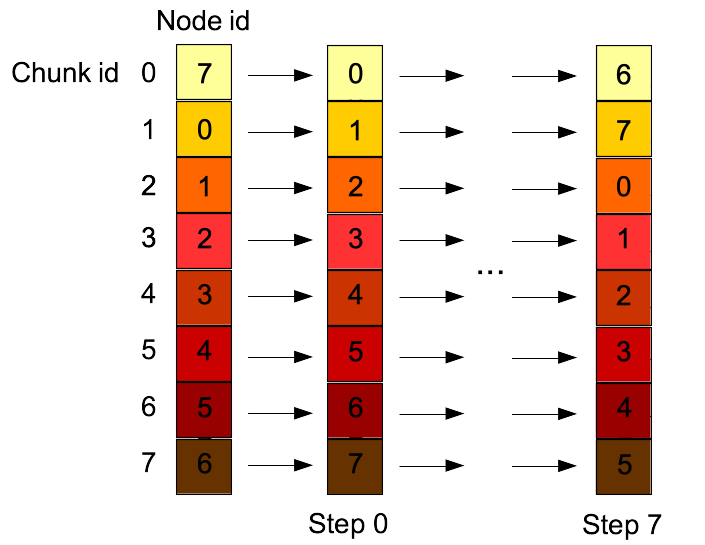}
\caption{Segmented pipelined ring Allreduce: Allgather stage where the partial results are broadcasted to the other nodes.}
\label{fig:allreduce_broadcast}
\end{figure}
The segmented pipelined ring algorithm consists of two stages: Scatter-Reduce and Allgather. On each of these stages, a process operates with 1/P of data, sending to the next clockwise node in the ring and receiving from the previous one in the ring.
The Scatter-Reduce stage works as follows: in the $k$th step, node $i$ will send the $i - k$th chunk and receive the $i - k - 1$th chunk, reducing it into its existing data of that chunk. Hence, each of the P nodes performs a reduction of 1/P of the dataset (see~\Cref{fig:allreduce_reduce}) and sends the result further. At the end of this stage, each node holds a complete result of 1/P of the data; we color this result on the plot. 

In the Allgather stage,
the fully accumulated partial results are distributed across all nodes, following again the pipelined ring communication with P-1 steps, as depicted in~\Cref{fig:allreduce_broadcast}. At the $k$th step, node $i$ will send chunk $i - k + 1$ and receive chunk $i - k$. 
After the Allgather, all nodes have access to the complete reduced dataset. 

A benefit of the segmented pipelined ring algorithm is that at every stage of Allreduce each process (often) deals with its close neighbors: receiving the partial data from one and sending the partial data to another.
%
Depending on the message size, we can also require a subspliting of messages: 1/P of the data can be divided into smaller messages to better utilize the network. As we rely upon GASPI API and not, for example, on ibverbs (which are used within GPI-2), we leave the splitting of messages to GPI-2 since it already handles this very efficiently. Hence, the GASPI implementation of Allreduce manages to use the entire memory and network bandwidth of the system. For the reduction, we used a global sum. Thus, we can hide the complete reduction effort in the communication costs. As long as the reduction effort is less time-consuming than the corresponding communication, this will also hold true for more complex reductions like user-defined reductions on user-defined data structures. 

Currently, we work on extending Allreduce towards eventually consistent collectives by coupling it with a compression technique. Hence, we foresee to reduce the amount of data transferred as well as to crop some data. 

\subsection{AlltoAll}
In a crucial part of the Quantum Espresso~\cite{QE-2009} application, the communication time is dominated by 
the {\tt MPI\_AlltoAll} collective used within a customized implementation of the Fast Fourier Transformation (FFT). In particular, {\tt MPI\_AlltoAll} consumes roughly 20-40\% of FFT’s total runtime.

To address this, we propose a preliminary design for an algorithmic variant of the AlltoAll collective leveraging the GASPI API. 
The underlying idea of this solution is to let each node write its data to the memory of all other nodes using {\tt gaspi\_write\_notify} with a unique notification. Then, each node waits on the notification ({\tt gaspi\_notify\_waitsome}) that some data has been written to its memory and resets this notification ({\tt gaspi\_notify\_reset}). After each node has written to all nodes and has received the data and notifications from them, the AlltoAll is complete.

\section{Experimental Results}
\label{sec:results}
We conducted our performance measurements on a set of different clusters:
\begin{itemize}
    \item SkyLake partition at Fraunhofer with a dual Intel Xeon Gold 6132 CPU @2.6\,GHz and 192\,GB of memory. Nodes are connected with the 54\,Gbit/s FDR Infiniband.
    \item Marenostrum4 cluster at BSC. Each node contains two 24-cores Intel Xeon Platinum 8160 CPUs @2.10\,GHz and 96\,GB of memory. Nodes are interconnected with the 100\,Gbit/s Intel OmniPath HFI Silicon.    
    \item Galileo cluster at Cineca. Each one contains two 18-cores Intel Xeon E5-2697 v4 (Broadwell) CPUs @2.30\,GHz and 128\,GB of memory. Nodes are interconnected through the 100\,Gbit/s Intel OmniPath with OPA v10.6.
\end{itemize}
We assign one GASPI process per node (unless otherwise mentioned) in order to stress the communication. 

\paragraph{Eventually consistent Allreduce} 
To experiment with our {\tt allreduce\_SSP} implementation, we use a simple  Matrix Factorization algorithm using Stochastic Gradient Descent (SGD), similar to~\cite{MFSGD}. The goal of this experiment is to understand the effect of setting different values of slack on the execution time of the algorithm and its convergence.
To train the model, we use the MovieLens 25M Dataset, iterate for a total of 500 iterations for the $slack=0$ execution, and then for the other executions with different values of slack use a number of iterations  necessary to achieve the same error.

\begin{figure*}[!ht]
\centering
\includegraphics[width=0.85\textwidth]{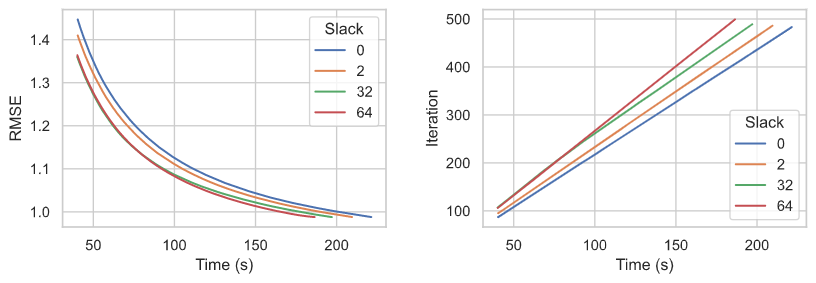}
\caption{Experimental results of {\tt allreduce\_SSP} impact on convergence speed of the ML algorithm (the Matrix Factorization algorithm using Stochastic Gradient Descent) 
 on 32 nodes of the MareNostrum4 cluster.}
\label{fig:allreduce_ssp_conv_speed}
\end{figure*}

\begin{figure*}[!ht]
\centering
\includegraphics[width=0.9\textwidth]{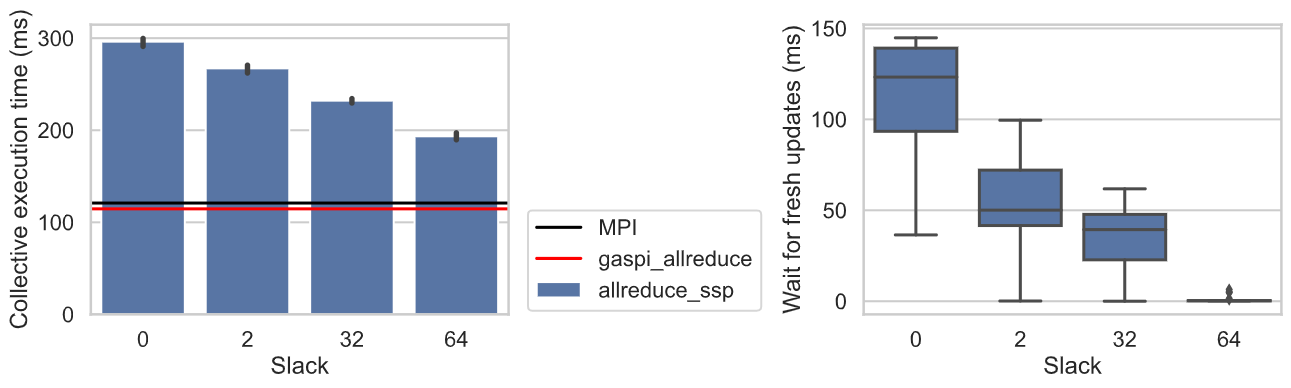}
\caption{Experimental results of {\tt allreduce\_SSP} collective execution speed and time spent waiting for fresh updates on 32 nodes of the MareNostrum4 cluster.}
\label{fig:allreduce_ssp_execution}
\end{figure*}

\Cref{fig:allreduce_ssp_conv_speed} shows the performance results of the {\tt allreduce\_SSP} evaluation using 32 GASPI processes spread across 32 nodes on MareNostrum4. The plotted points were sampled once per iteration and correspond to the average (per iteration) of all 32 workers.

The plot on the right shows the behavior of the iteration speed with respect to different values of slack. This highlights that, after a given execution time (i.e., when fixing the value on the $x$ axis), more iterations are performed as we increase slack.  This happens because of the fact that the SSP condition for advancing from one iteration to the next is more relaxed than in the traditional execution ($slack=0$), allowing workers to lag up to slack iterations behind. 

When benefiting from slack, faster processes are satisfied, on average, with potentially increasingly staler data in order to avoid waiting for new contributions. Because of this, at a certain point in their execution, processes will try to use contributions that are staler than their allowed slack. 
To illustrate this, we focus on the behavior of the execution using $slack=32$ and $slack=64$. Up to around the 100th second on the plot on the right hand side, both executions are rather similar. However, after this point we see that the $slack=64$ execution maintains its iteration per second, while the execution with $slack=32$ decreases its iterations per second. 
At this point, processes using $slack=32$ are trying to use contributions that are staler than 32 clocks, while the execution using $slack=64$ can continue to use staler data. Hence, this is reflected by the decrease in iteration per second in the $slack=32$ execution.


The fact that iterations run faster may not imply that the convergence time improves since the algorithm will take more iterations to converge.
However, the plot on the left in~\Cref{fig:allreduce_ssp_conv_speed} shows that there is an overall gain in the execution time. In particular, it shows that, for algorithms with a convergent nature such as Matrix Factorization using SGD, the Allreduce implementation following the SSP approach can increase the overall convergence speed by reaching the desired error faster.
Compared to $slack=0$, $slack=2$ required 3 more iterations to reach the same error while being 6\% faster, $slack=32$ required 6 more iterations and was 12.3\% faster, $slack=64$ required 16 more iterations and was 19\% faster.

In~\Cref{fig:allreduce_ssp_execution}, on the left plot, we see the collective execution time for the {\tt allreduce\_SSP} solution. From the plot, we can see that this solution performs significantly worse than the collective offered by MPI (default pick of Allreduce) and the {\tt gaspi\_allreduce\_ring} implementation. Note that both {\tt gaspi\_allreduce\_ring} and MPI use the Allreduce algorithms more suited for large vectors.
In fact, even in the configuration for the slack value with the lowest execution time, this solution is still around 58\% slower than our baseline collectives. 
This is due to the fact that the Hypercube algorithm shuffles the entire vector, which is better suited for small vectors, and we use vectors of a considerable size.

Regardless of the absolute performance, we see that the {\tt allreduce\_SSP} collective benefits from higher values of slack by being able to reduce, and even completely eliminate, the time waiting for fresh updates as shown in the right plot of~\Cref{fig:allreduce_ssp_execution}. 
As mentioned earlier, we did not expect this solution to have stellar performance. Instead, we designed it as a first attempt at adapting an existing step based Allreduce algorithm to use SSP and determine if SSP would reduce their execution time, which we confirmed.



\begin{figure*}
\begin{center}
\begin{tabular}{cc}
\hspace*{-4mm}\begin{minipage}[t]{0.48\textwidth}
\includegraphics[width=1.02\columnwidth]{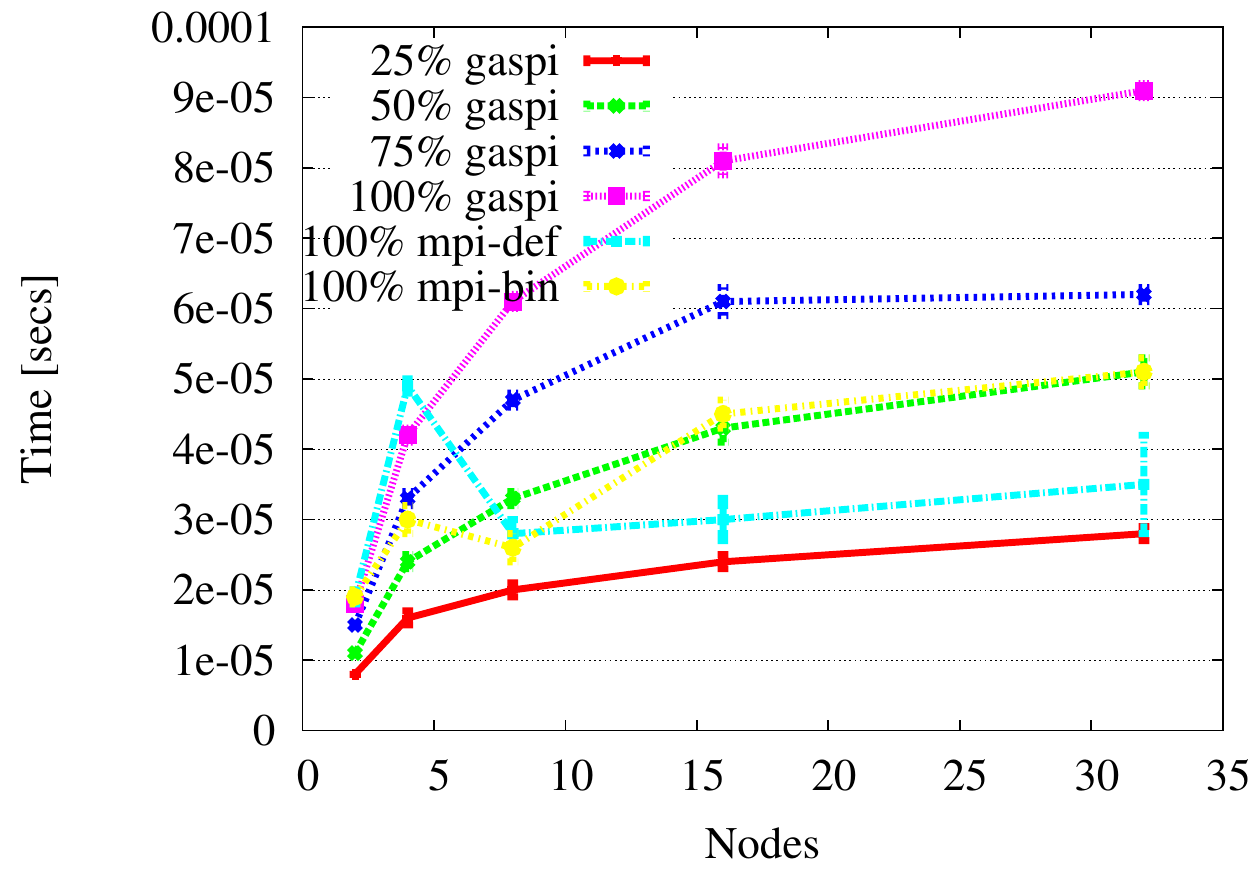}
\end{minipage}
&
\begin{minipage}[t]{0.48\textwidth}
\includegraphics[width=1.02\columnwidth]{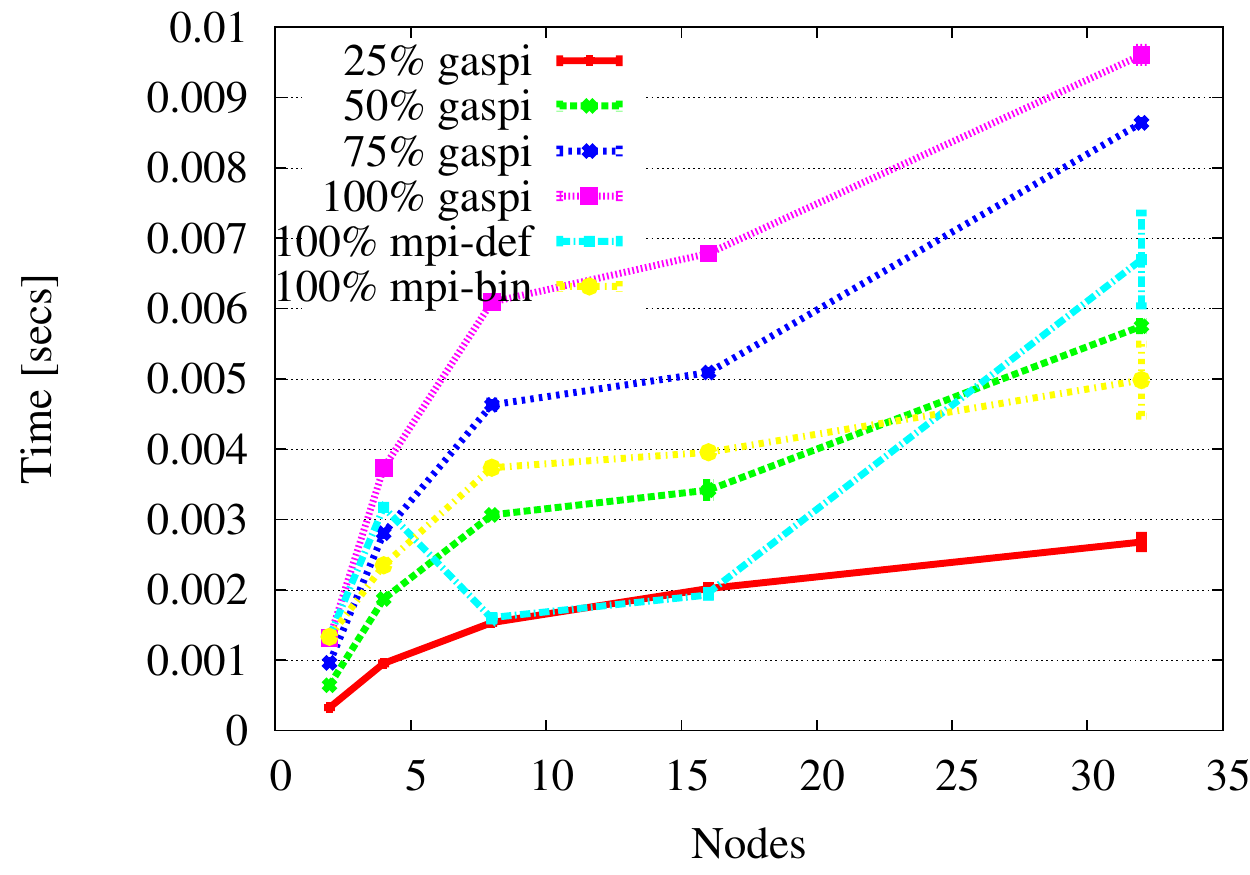}
\end{minipage}
\end{tabular}
\vspace{-10pt}
\caption{\label{fig:bcast} Performance results of Broadcast on SkyLake nodes: for vectors of 10,000 double precision elements on the left and of 1,000,000 elements on the right. mpi-def stands for the default variant of Broadcast, while mpi-bin corresponds to the binomial variant.}
\end{center}
\end{figure*}
\begin{figure*}
\vspace*{-2.5mm}
\begin{center}
\begin{tabular}{cc}
\hspace*{-4mm}\begin{minipage}[t]{0.48\textwidth}
\includegraphics[width=1.02\columnwidth]{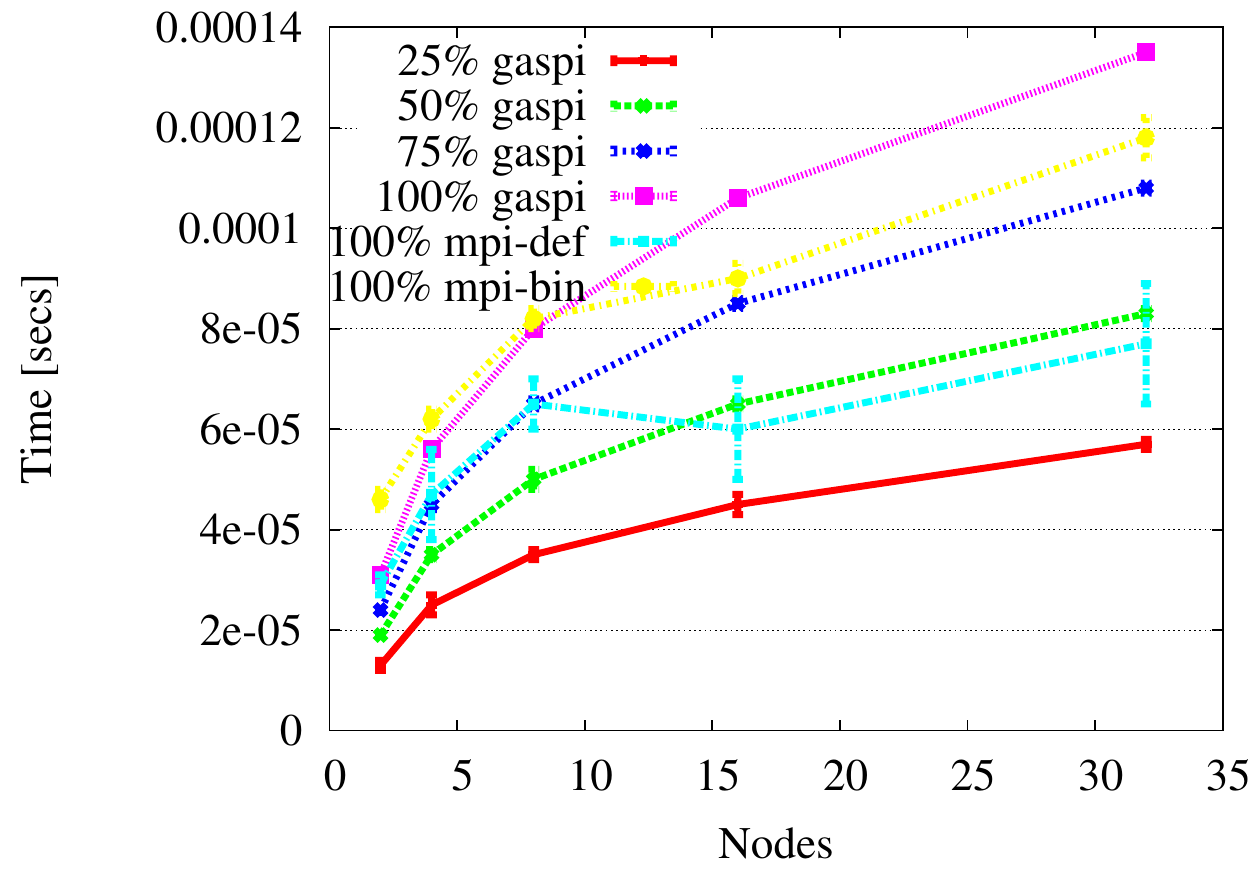}
\end{minipage}
&
\begin{minipage}[t]{0.48\textwidth}
\includegraphics[width=1.02\columnwidth]{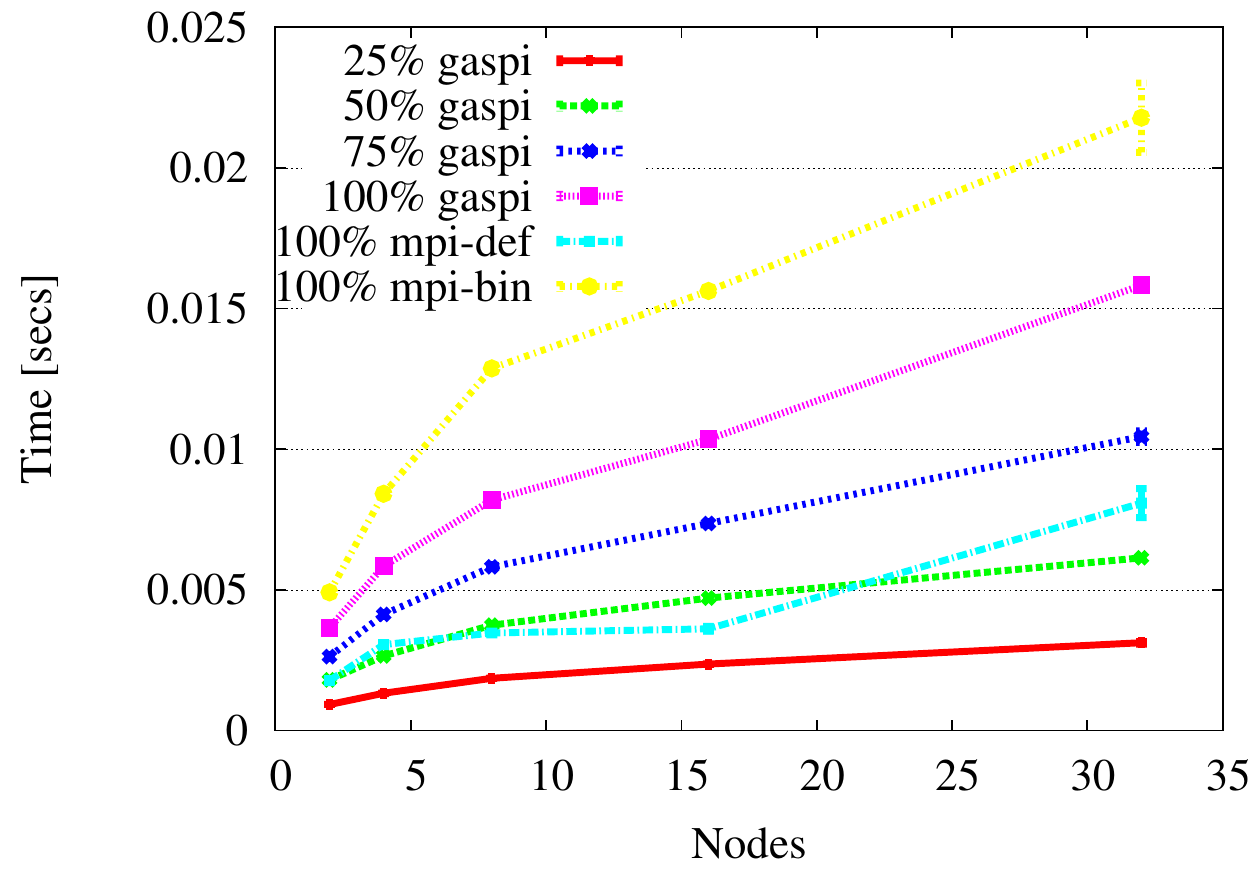}
\end{minipage}
\end{tabular}
\vspace{-10pt}
\caption{\label{fig:reduce} Performance results of Reduce on SkyLake nodes: for vectors of 10,000 double precision elements on the left and of 1,000,000 elements on the right. mpi-def stands for the default variant of Reduce, while mpi-bin corresponds to the binomial reduction.}
\end{center}
\vspace*{-2.5mm}
\end{figure*}
\begin{figure}[!ht]
\centering
\vspace*{-2.mm}
\includegraphics[width=1.02\columnwidth]{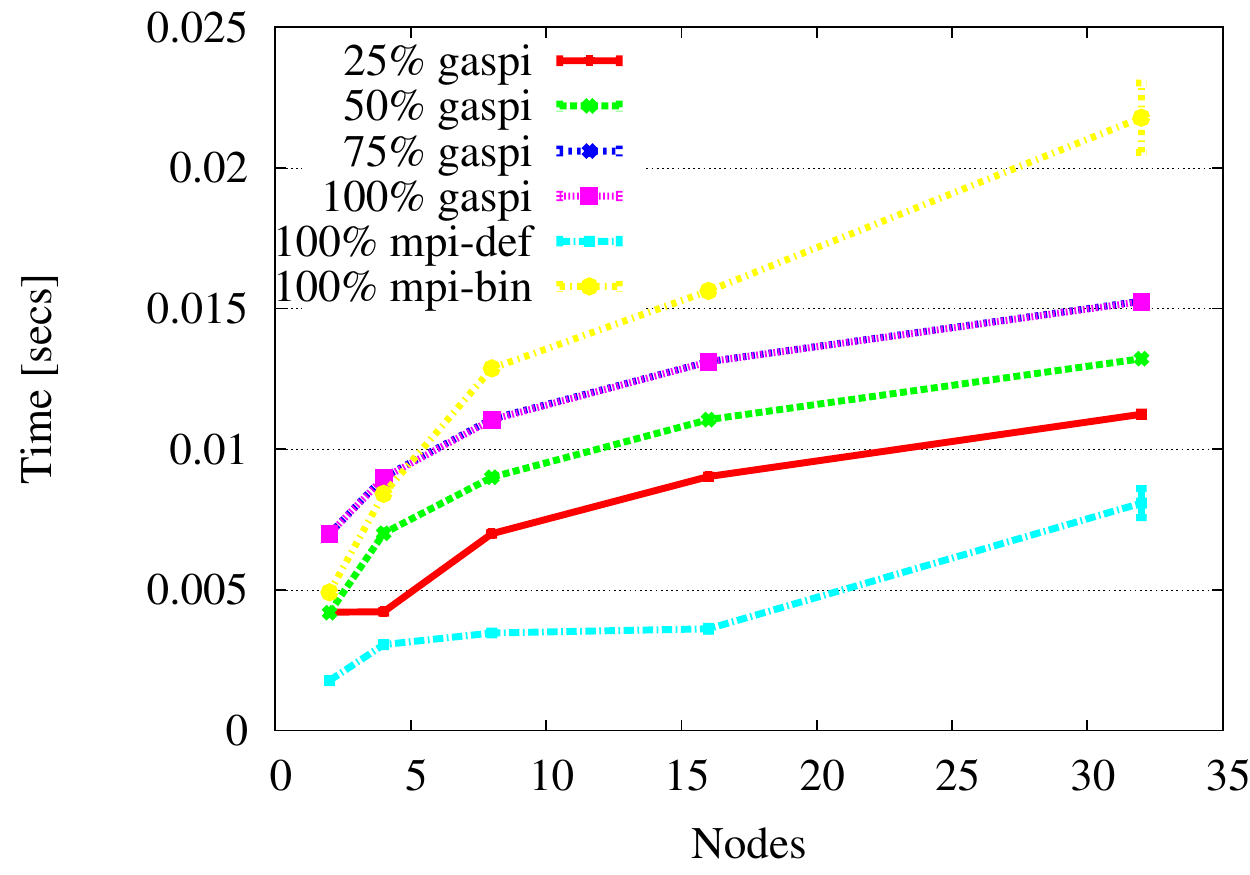}
\caption{Performance results of Reduce operating on the full amount of data for 1,000,000 doubles on SkyLake nodes; xx\,\% lines corresponds to at least xx\,\% of processes involved.}
\label{fig:reduce_full}
\vspace*{-3.5mm}
\end{figure}

\paragraph{Eventually consistent Broadcast}
Broadcast is often used in HPC and ML/ DL code to distribute the initial data for computations. Hence, its impact on the overall application performance can be very small. In cases when Broadcast is also used within applications, especially on every iteration, careful balancing between the most relevant data and its amount can lead to significant savings in terms of the execution time. \Cref{fig:bcast} shows the performance results of Broadcast using different thresholds for data, i.e., different amounts of data that is shipped.  We conduct 100 executions for each message size and calculate the average time among all executions. We also compute the 95\,\% confidence interval that is displayed as error lines on the plots. The GASPI BST variant of Broadcast is 3.25x-3.58x faster when dealing with only a quarter of the data. Moreover, we compare these performance results against the ones with {\tt MPI\_Bcast} from Intel MPI version 2018 update 1: {\tt mpi-bin} is the {\tt MPI\_Bcast} implementation with a binomial tree; {\tt mpi-def} is the default (automatically selected) broadcast implementation. These two MPI variants are clearly better compared to the {\tt gaspi\_bcast} on small data sets (up to few thousands of elements), possibly using a different binomial implementation. However, the overhead of our implementation decreases on a larger node count and for large arrays, where we can see promising performance benefits of {\tt gaspi\_bcast}. As such, we are considering to revise the BST implementation as well as to implement an alternative variant. 

\paragraph{Eventually consistent Reduce}
We conduct similar tests for Reduce and report the respective results in~\Cref{fig:reduce}. 
For each message size, we also run the benchmark 100 times and compute both the average and the 95\,\% confidence interval. 
For different message sizes, the difference between the usage of 25\,\% and 100\,\% of the data in {\tt gaspi\_reduce} increases rapidly with the message size and for 8\,Mb it is roughly 5x. 
We compare the performance results of {\tt gaspi\_reduce} against the ones with {\tt MPI\_Reduce} from Intel MPI version 2018 update 1 using their default (automatically selected) and binomial variants. For small arrays, MPI outperforms all our variants of Reduce. However, for larger arrays, starting from 10,000 elements, the situation changes: the automatically selected variant (from the pool of 14) of {\tt MPI\_Reduce} is still (1.96x) faster, while {\tt gaspi\_reduce} is roughly by 38\,\% faster than the MPI binomial variant.

We also provide performance results of another {\tt gaspi\_reduce} implementation when the full amount of data is sent but only a certain (according to the pre-defined threshold) percentage of processes is engaged in communication, see~\Cref{fig:reduce_full}. Hence, the leaves that are far from the root are excluded. This implementation  has slower performance compared to the previous implementation of {\tt gaspi\_reduce}, working only on a fraction of data, however it is still better then the MPI binomial variant. The lines for 75\,\% and 100\,\% show identical performance due to the fact that after 50\,\% it is difficult to get much improvement since 50\,\% of all processes are added on the last stage of the algorithm as depicted in~\Cref{fig:bst}. 

\begin{figure*}
\begin{center}
\begin{tabular}{cc}
\hspace*{-4mm}\begin{minipage}[t]{0.48\textwidth}
\includegraphics[width=1.02\columnwidth]{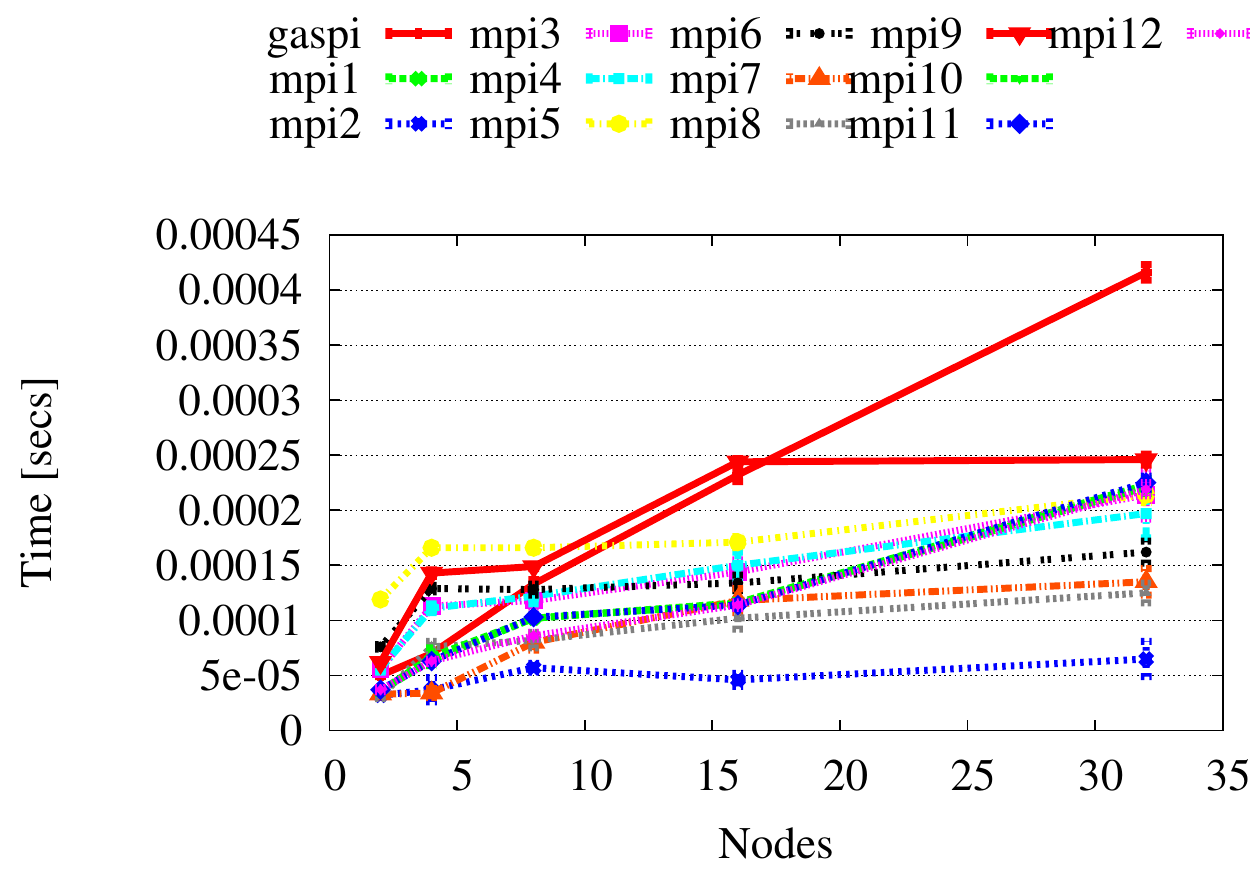}
\end{minipage}
&
\begin{minipage}[t]{0.48\textwidth}
\includegraphics[width=1.02\columnwidth]{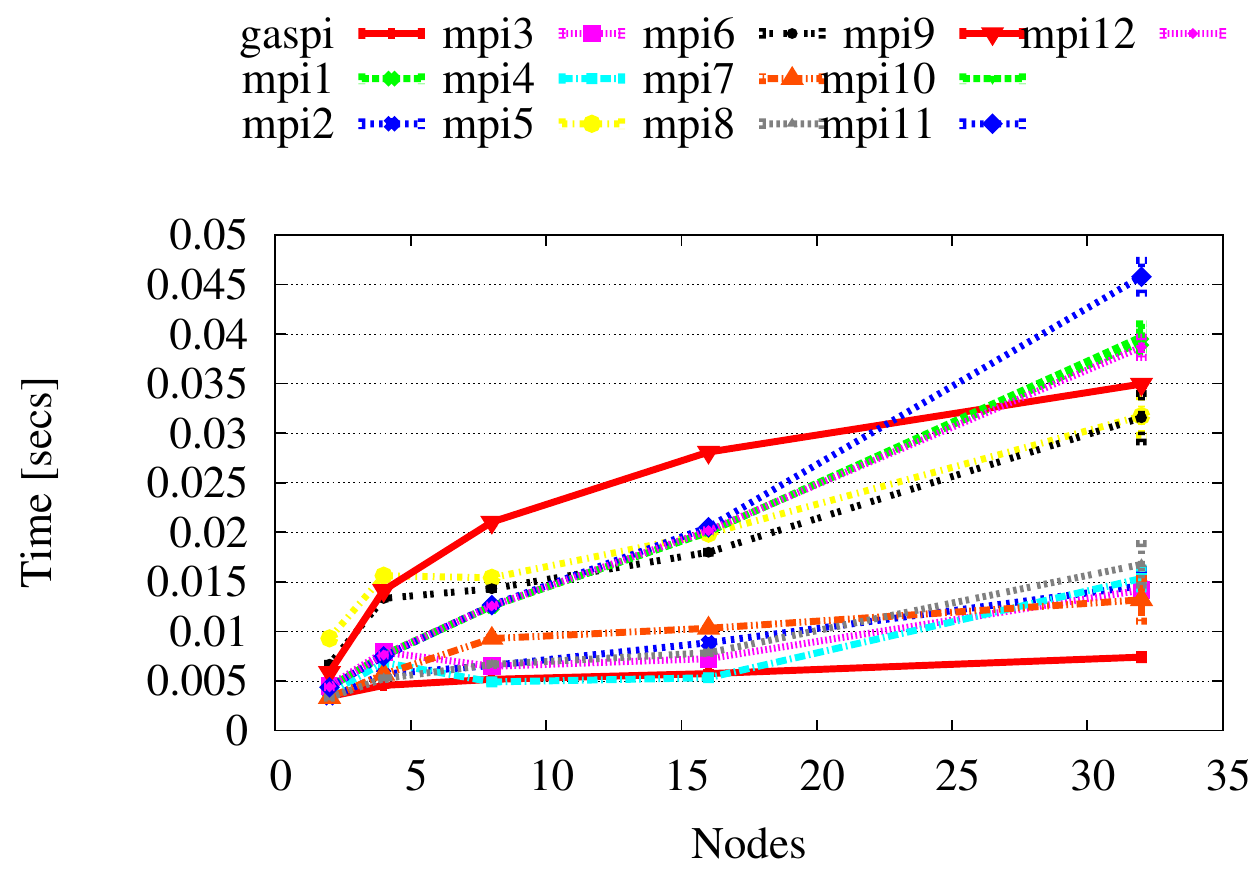}
\end{minipage}
\end{tabular}
\caption{\label{fig:allreduce} Performance results of Allreduce on SkyLake nodes: for  vectors of 10,000 double precision elements on the left and of for 1,000,000 elements on the right. gaspi corresponds to the segmented pipelined ring with GASPI ({\tt gaspi\_allreduce\_ring}); mpi1 stands for the recursive doubling variant; mpi2 -- Rabenseifner's; mpi3 -- Reduce + Bcast; mpi4 -- topology aware Reduce + Bcast; mpi5 -- binomial gather + scatter; mpi6 -- topology aware binominal gather + scatter; mpi7 -- Shumilin's ring; mpi8 -- ring; mpi9 -- Knomial; mpi10 -- topology aware SHM-based flat; mpi11 -- topology aware SHM-based Knomial; mpi12 -- topology aware SHM-based Knary.}
\end{center}
\end{figure*}
\begin{figure}[th!]
\begin{center}
\hspace*{-4mm}\includegraphics[width=1.04\columnwidth]{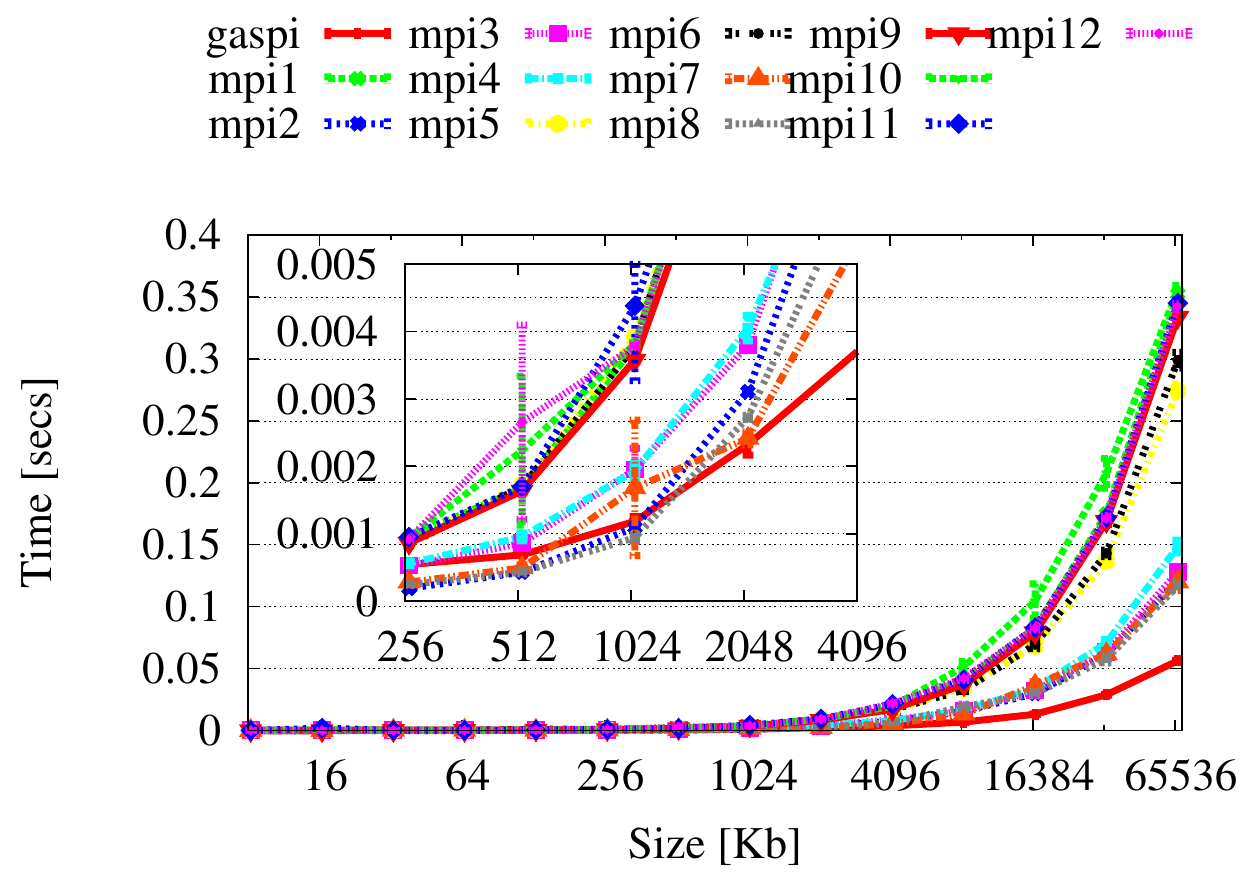}
\caption{\label{fig:allreduce:more} Performance results of Allreduce on 32 SkyLake nodes for various message sizes.}
\end{center}
\end{figure}
\begin{figure}[!ht]
\begin{center}
\includegraphics[width=\columnwidth]{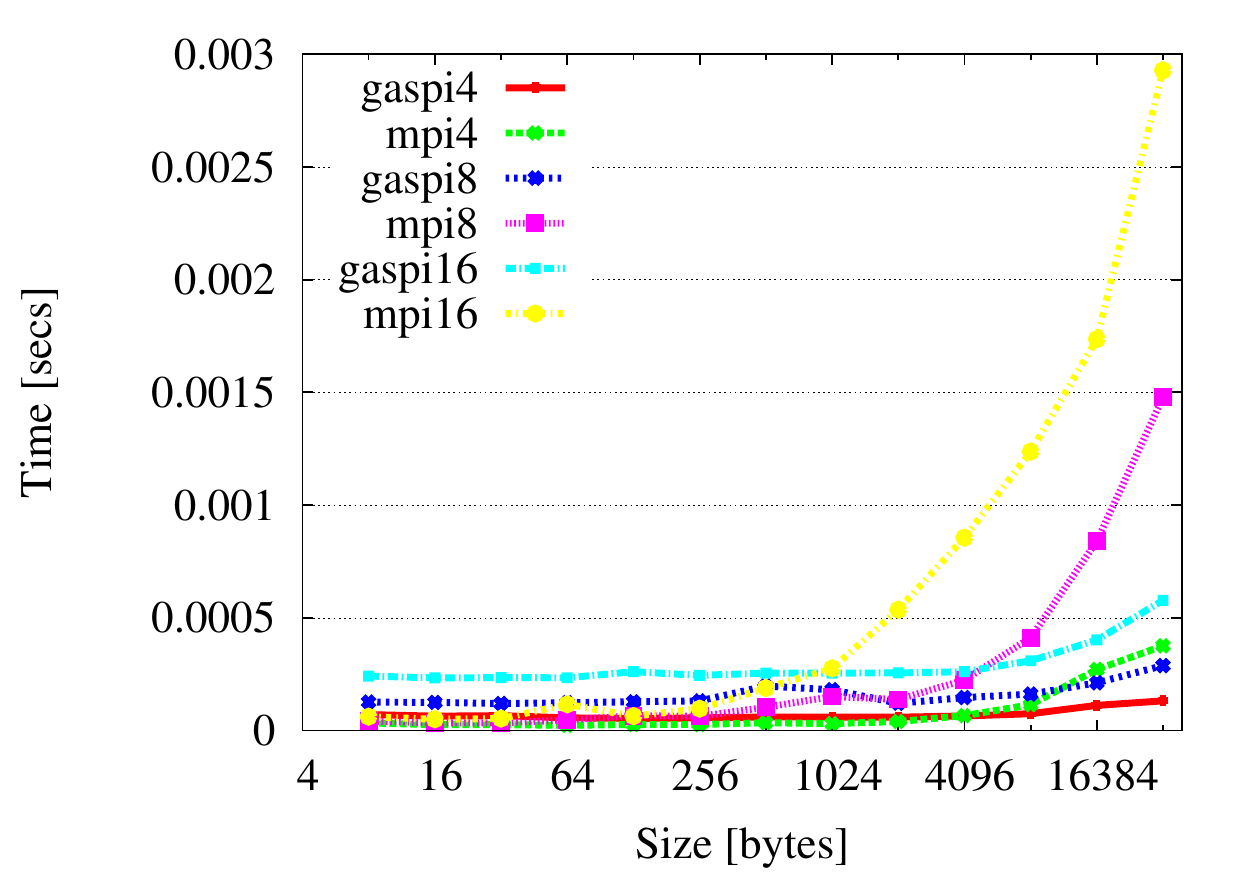}
\caption{\label{fig:alltoall} Performance results of {\tt gaspi\_alltoall} compared against MPI on the Galileo cluster at CINECA; GPI-2 installation is from the next branch on GitHub, which is v1.4.0, and MPI implementation comes from the Intel MPI v18.0 library.}
\end{center}
\end{figure}

\paragraph{Classic/ Consistent Allreduce}
We carry out the experiments of the segmented pipelined ring Allreduce with GASPI (called {\tt gaspi\_allreduce\_ring}) on the SkyLake nodes of the Fraunhofer ITWM's cluster. We compare its performance results against the ones of {\tt MPI\_Allreduce} from Intel MPI version 2018 update 1 with the standard settings, which ensures that the optimal settings are being selected.  
We allocated one process per node and assigned the same amount of work (vector size) per node in order to stress communication of the collective. \Cref{fig:allreduce} reports the timings (the average time and the 95\,\% confidence level among 100 executions) for arrays of doubles of sizes 10,000 and 1,000,000. Note that our implementation aims to target large vector sizes, whereas the Intel MPI library is equipped with a dozen of implementations, which in fact are all in use here, targeting various message sizes and topologies. Hence, the results of {\tt MPI\_Allreduce} are significantly better than {\tt gaspi\_allreduce\_ring} for vectors of size 10,000. However, {\tt gaspi\_allreduce\_ring} performs better for larger vectors, e.g., with 1,000,000 elements, showing performance benefits of 1.78x and 2.26x when compared against the Shumilin's ring (mpi7 on the plot) and the ring (mpi8) variants of {\tt MPI\_Allreduce}, respectively. 
 The Shumilin’s ring is the best performing variant of MPI among the existing 12. The ring MPI variant supposedly implements the same segmented pipelined ring algorithm. 
It is worth mentioning that, compared to the implementations of the segmented pipelined Allreduce with MPI, we eliminate global synchronizations at the end of both Scatter-Reduce and Allgather phases and instead use the GASPI weak lightweight synchronization via notifications; for example, to indicate that the process is ready to receive data or to acknowledge arrival of data as depicted in~\Cref{fig:gaspi:write}.

In addition, we conduct tests by focusing on a range of message sizes, for instance from 1,024 elements and up to eight millions with the step of 2. \Cref{fig:allreduce:more} demonstrates the results of this evaluation: 1) until we reach a message size of 1,048\,Kb, the MPI implementation of Allreduce is faster; 2) starting from a message size of 2,097,Kb our implementation outperforms all MPI variants and reaches its peak of 2.07x and 2.13x for ring and Shumilin's ring variants, respectively, for a message size of 67,108\,Kb (or 8,388,608 elements). We believe that this trend will continue and the gap will become larger on more nodes since the MPI allreduce curves show faster growth. This is particularly thanks to the GASPI's asynchronous one-side communication with weak synchronization.

\paragraph{Classic/ Consistent AlltoAll}
We conducted our performance experiments and compared the GASPI implementation of AlltoAll against that of {\tt MPI\_AlltoAll} of Intel MPI v.18.0 on Cineca's Galileo cluster. 

\Cref{fig:alltoall} reports the performance results (obtained as an average over 100 runs) for various message sizes. Since we aim to have a hybrid programming model implementation, we set four GASPI/ MPI processes per node. 
We run our experiments using 4, 8, and 16 nodes: this is marked on each line as gaspi$x$ or mpi$x$, where $x=4,8, \text{or } 16$. Note that we used Intel MPI v.18.0 with the standard settings, which ensures that the optimal settings are being selected. The performance of both implementations is similar up to a message size of about 1,024 bytes. The situation begins to change from a message size of 2,048 bytes, where our implementation of GASPI AlltoAll begins to outperform the vendor-provided MPI  version, reaching a peak performance for a message size of 32,768 bytes; the performance gain is 2.85x, 5.14x, and 5.07x on 4, 8, and 16 nodes, respectively.

It is important to note that the message size needed by the FFT miniapp when using {\tt MPI\_AlltoAll} is in the range of 6Kb-24Kb, i.e., in the range where the GASPI version outperforms the vendor-provided MPI implementation. Since {\tt MPI\_AlltoAll} consumes about 20-40\% of FFT’s total runtime, we expect a significant reduction of the total execution time in the Quantum Espresso application (whose implementation is currently in progress). A GASPI equivalent of {\tt MPI\_AlltoAllV}, also used under certain conditions in the minapp, is currently being built using the same scheme as for the {\tt gaspi\_alltoall} collective. 

\section{Related Work}
\label{sec:relatedworks}

Depending on message sizes and network architecture, Allreduce implementations span a wide range of algorithms, from ring-based algorithms, binomial spanning-tree implementations~\cite{tzeng1998fast},  tree algorithms~\cite{mellor1991algorithms}, or Butterfly like algorithms~\cite{brooks1986butterfly}. In~\cite{end2015adaption} the authors designed an n-way dissemination algorithm for the GASPI API; this algorithm is suitable for small message sizes. 
The  dissemination algorithm has been presented by Hensgen  et  al. in 1988~\cite{hensgen1988two}.  Due to its speed it is  used  in  different  programming  APIs and  libraries  like  the  MPICH  implementation of  MPI for  barrier  implementations. 
In~\cite{end2015adaption}, the n-way dissemination algorithm was used in a way that neither leverages partial reductions of a 2-way dissemination (and their associated out-of-band delivery to late dissemination stages), nor notified communication in shared windows.

The GASPI programming model~\cite{gaspi15} primarily targets multi-threaded or task-based  applications, hence GASPI+X. However, in  order  to  support migration   of   legacy   applications   (with   a   flat   MPI communication model) towards GASPI, we have extended the  concept  of  shared  MPI  windows~\cite{hoefler2013mpi+,gropp2014using} towards  a  notified  communication model~\cite{PMAA-MPI-GASPI,IJHPCA-MPI-GASPI} in  which the  processes  sharing  a  common  window  become  able  to see all one-sided and notified communication targeted at this window. 
 We enabled communication from and to a shared memory region to all processes, which share the window. 
While MPI-3 readily supports this model with MPI 2-sided and 1-sided communication, we aimed to support the notified and one-sided communication in GASPI. Since GASPI does not require a dedicated receiving process, we can avoid the detrimental effects of late receivers. Nevertheless, all processes are still able to test for completion of incoming messages without additional synchronization effort.

In~\cite{PMAA-MPI-GASPI,IJHPCA-MPI-GASPI} we also designed Allreduce based on pipelined rings and notified communication in shared windows. This implementation delivered up to the 3x performance boost compared to the best Intel MPI implementations v5.1.2 on the Salomon IT4I cluster (Infiniband FDR). 
We extended this idea to Allgather(V) and Allreduce with an adaptation of the dissemination algorithm~\cite{Iakymchuk19GASPIcoll}, achieving up to 2x-4x performance improvements compared to the best performing MPI implementations on the Salomon IT4I cluster and the Beskow Cray XC40 cluster at PDC, KTH. 

\section{Conclusions and Future Work}
\label{sec:conclusions}
In this article, we presented our ideas for adapting some classic algorithm for collective operations -- like Binomial Spanning Tree and segmented pipelined ring -- to implement Broadcast, Reduce, and Allreduce with GASPI. While with Broadcast and Reduce we aimed to be generic but vary the amount of data used or processes involved (mimicking eventual consistency), with Allreduce we targeted (very) large message sizes. The Allreduce implementation leads to 2x faster execution compared against a dozen of the vendor-specific implementations. We also implemented AlltoAll following a rather simple but well-performing pattern, resulting in 2.8x-5.1x performance improvements compared to MPI's AlltoAll default variant. 

Furthermore, we designed and implemented in GASPI a novel Allreduce following the Stale Synchronous Parallel model ({\tt allreduce\_ssp}). 
{\tt allreduce\_ssp} reduces the waiting time and synchronizations by using stale contributions. This approach is suitable for ML/ DL computations. 
Our implementation, which is based on Hypercube, was not able to outperform the MPI standard, however we observed the desired effect of using slack in terms of faster convergence for Matrix Factorization. In order to improve {\tt allreduce\_ssp}, we consider to adapt more efficient Allreduce algorithms, e.g. the presented pipeline ring algorithm, and to explore the idea of the Parameter Server architecture, which is the setting where we usually find the SSP model.
All our developments are available under the EPEEC GitHub repository: \url{https://github.com/epeec}.

Our ultimate goal is to provide a library of collectives within the GASPI ecosystem for both the HPC and ML/ DL communities by leveraging the GASPI API and focusing on the design of collectives for various data sizes and/ or application needs as in case of the {\tt allreduce\_ssp} variant. Furthermore, we are also working on the compression library and foresee to design and develop another version of eventually consistent collectives by coupling this compression library with the consistent collectives.  

\section*{Acknowledgement}
The research leading to these results has received funding from the European Union's Horizon2020 research and innovation programme under the EPEEC project, grant agreement No 801051, and from FCT under UIDB/50021/2020.

\bibliographystyle{IEEEtran}
\bibliography{main}

\end{document}